\documentclass{amsart}
\RequirePackage{fix-cm}

\usepackage{wasysym}
\usepackage{amssymb,amsmath,stmaryrd,amsfonts}
\usepackage{mathtools,leftindex}
\usepackage{multirow}

\usepackage{array}

\usepackage{tikz,mwe,pifont}
\tikzset{boximg/.style={remember picture,red,thick,draw,inner sep=0pt,outer sep=0pt}}

\usepackage{natbib}
\usepackage[margin=1.2in]{geometry}

\newcommand{\ie}{\textit{i.e.,}}
\newcommand{\eg}{\textit{e.g.,}}

\newcommand{\norm}[1]{\left\lVert#1\right\rVert}

\newcommand{\tred}[1]{\textcolor{black}{#1}}

\usepackage{enumitem}   

\usepackage{rotating}

\usepackage[colorlinks,citecolor=blue]{hyperref}

\begin{document}

\title[Divergence of RPs and chaos]{Leveraging temporal features of the divergence quantifier of recurrence plot to detect
	chaos in conservative systems}

\author[J.\,Daquin]{J\'er\^ome Daquin}
\address{
	Universit\'e de Toulon, Aix Marseille Univ, CNRS, CPT, Toulon, France
	}
\email{jerome.daquin@cpt.univ-mrs.fr}

\author[T.\,Kov\'acs]{Tam\'as Kov\'acs}
\address{
	Institute of Physics and Astronomy, E\"otv\"os University, Budapest, Hungary
}
\email{tamas.kovacs@ttk.elte.hu}

\date{\today}

\begin{abstract}
The recurrence-based divergence quantifier ($DIV$), traditionally applied to dissipative systems, is shown here to be an effective finite-time chaos indicator for conservative dynamics. We benchmark its performances against the well-established fast Lyapunov indicator (FLI), focusing on the standard map, a canonical model of Hamiltonian chaos. Through extensive numerical simulations on moderately long orbits, we find strong agreement between $DIV$ and FLI, supporting the reported correlation between the divergence of recurrences and positive Lyapunov exponents. Additionally, 
our study sheds more light into asymptotic time properties of $DIV$  by revealing distinct power laws on regular and chaotic components, both in the original and reconstructed phase spaces. In particular, on a regular component, the space average of $DIV$ decays  with the time $N$ as $1/N$, mirroring the decay rate of the maximal Lyapunov exponent.  \tred{On chaotic components},
\tred{the space average of} $DIV$ decreases at a much slower rate, close to $1/\sqrt{N}$.
This scaling insight opens new avenues for characterizing chaos from time series.
Our numerical results thus demonstrate $DIV$ to be a computationally viable and theoretically rich tool for chaos detection in conservative systems.
\end{abstract}


\maketitle

\tableofcontents

\section{Introduction}
The story of recurrence plot (RP) and the beginning of its methodological developments is often traced back to 
the seminal paper of \cite{jpEc87}.
Let be $\gamma$ a finite orbit of points in $\mathbb{R}^{d}$, $d \ge 1$,
$\gamma=\{z_{0},z_{1},z_{2},\cdots,z_{n}\}$,  of some dynamical systems. 
The $\{z_{i}\}_{i}$ might be the results of mapping iterations, flow discretisation or observed states (measurements). 
The $0-1$  recurrence matrix $R_{\gamma}=
(r_{i,j})_{0 \le i,j \le n}$
is constructed
from the recurrence scalar 
\begin{align}\label{eq:rijRP}
	r_{i,j} = \Theta\big(\tred{\epsilon_{i}}-\norm{z_{i}-z_{j}}\big), 
\end{align}
where $\Theta$ is the Heaviside function,  $\tred{\epsilon_{i}}$ is a real cutoff parameter, and 
$\norm{\bullet}$ is a norm in $\mathbb{R}^{d}$. 
When $r_{i,j}=1$, $z_{i}$ and $z_{j}$ are recurrence points, which means that they are $\tred{\epsilon_{i}}$ close. The matrix $R_{\gamma}$ always contains the line of identity as trivial recurrence points \tred{and tends to be farily symmetrical\footnote{\tred{When the cutoff parameter $\epsilon$ is independent of the time indices $i$, the recurrence plot is symmetric with respect to the first  diagonal.}}}. 
A RP, as indicated by the name, is in its root a graphical tool. It  consists of visualising the binary matrix $R_{\gamma}$ with recurrence points encoded as dark pixels. Interestingly enough, dynamical properties of the orbit (such as its oscillatory nature, the presence of drifting components, the presence of extreme or rare events, etc.) are encoded in different structures and textures of the RP (\cite{jpEc87,chWe94}). 
(Several typologies obtained on the standard map, the model on which we base our analysis, will be exemplified later in Sec.\,\ref{sec:methods}.) 
In addition to qualitative graphical assessments of RP, which can be difficult to interpret or intuit visually, several recurrence variables have been introduced by \cite{chWe94} and developed in greater depth over the years  \citep{nMa07-PR}, a field termed recurrence quantification analysis (RQA).   
Such variables include the recurrence rate $RR$ (percentage of recurrence points), the determinism $DET$ (proportion of recurring points forming diagonal lines) or lengths of vertical lines, to name but a few. \\

In this contribution, we focus specifically on the RQA variable related to the length of the longest diagonal lines $\ell_{\max}$  in a RP (leaving aside the line of identity and its vicinity), 
and more precisely its inverse --- another RQA variable --- the divergence $DIV$ \citep{chWe94}, defined as 
\begin{align}
	DIV=1/\ell_{\max}.
\end{align}
A diagonal line of $\ell$ length units emanating from the times $(i,j)$ in a RP corresponds to a sequence of $\ell$ successive recurrence points,  
\begin{align}
	\left\{
	\begin{aligned}
		&r_{i-1,j-1}=0, \, \\
		&r_{i,j}=1, \, \\
		&r_{i+1,j+1}=1, \\
		&\cdots  \\
		&r_{i+\ell-1,j+\ell-1}=1, \\
		&r_{i+\ell,j+\ell}=0. \, \\
	\end{aligned}
	\right.
\end{align}
It has long been reported $\ell_{\max}$ (or $DIV$) to be in direct proportion to the largest Lyapunov exponent \citep{jpEc87,jpZb92,chWe94,lTr96}, as heuristically understood here.  
Assume the trajectory to be embedded into some $\mathbb{R}^{d}$ and that $r_{i,j}$ is a recurrence point. It follows that
$z_{i}$ is $\tred{\epsilon_{i}}$ close to $z_{j}$. 
If we 
interpret $\delta_{i,j}=z_{i}-z_{j}$ as a deviation  vector and follow its time evolution, in case of chaotic dynamics, we expect the quantities 
$\{\norm{\delta_{i+k,j+k}}\}_{k}$ to grow exponentially fast in average with $k$ at a rate dictated by the largest Lyapunov exponent. Thus, the probability of 
$z_{i+k,j+k}$  \tred{to be still} an $\tred{\epsilon_{i}}$ neighbour of $z_{i,j}$ (a recurrent point) decreases with time $k$ very fast. 
In other words, the length of the diagonal that stems from  $(i,j)$ is expected to be small. \\

A more rigorous link between diagonal lines and chaos is made apparent when looking at the distribution of diagonal lines of length $\ell$ and the second-order R\'enyi entropy (also called correlation entropy), $K_{2}$ \citep{pGr83}. For purely deterministic systems $K_{2}=0$ and for stochastic systems $K_{2} \to +\infty$. Deterministic chaos is characterised by finite values, $K_{2} > 0$. It turns out that $K_{2}$ can be estimated directly from RPs as presented in \cite{mTh03}.  
An algorithm to allow its numerical estimation is detailled by \cite{nAs04} and \cite{nMa07-PR}. The procedure involves essentially three major steps. Firstly, one need to compute the cumulative distribution of the diagonal lines for several choice of $\epsilon$. Secondly, one need to identify automatically a scaling region and clusters with $\epsilon$. Finally, a last step involving a linear regression leads to the estimation of $K_{2}$. \\

Although  \tred{several} contributions have applied the RP \tred{framework} and RQA  \tred{quantifiers} to conservative dynamics  (see, \tred{for examples,} \cite{yZo07,yZo07-1,yZo16,mPa22,mrSa23}, or  \cite{nAs04,tKo19,tKo20} in the context of  gravitational $n$-bodies like dynamics), their developments have been mostly driven 
by dissipate dynamics. 
This might be partly due to the important application of RPs and RQAs to the field of time series analysis (see \cite{nMa23} for a bibliographical view of RPs and RQAs), the raise of chaos in low-dimensional dynamical systems, and the desire to analyse complex systems from the nonlinear time series perspective (intimately connected with phase space reconstruction and delay embedding theorem, see, \eg \, \cite{hAb93}). 
This contribution has two main objectives: first, to further apply the methodology and potential of RPs to conservative problems by focusing on the standard map model, paradigm of Hamiltonian chaos; and second, to propose a methodology for using the specific $DIV$  \tred{quantifier} as an indicator of chaos. 
Indeed, our work provides numerical evidences that the more heuristic divergence $DIV$ furnishes a simple and robust workaround of the more cumbersome automatisation of the estimation of $K_{2}$, and might be reliably used to distinguish between regular and chaotic motions. \\

The landscape of chaos indicators and complexity measures developed over the last decades is rich, and still an active field of research. Among the various existing methods, it is rather customary to distinguish between variational methods, relying on the Jacobian  associated to the dynamics (or, an estimation of it), and orbit based diagnostics. The divergence $DIV$ has the advantage to belong to the later family; yet, its complexity is $\mathcal{O}(N^{2})$, $N$ being the length of the orbit. 
A non-exhaustive list of indicators include the Lyapunov exponent \citep{gBe76},  
the fast Lyapunov indicator and variations of it \citep{cFr97,mF022}, the mean-exponential growth of nearby orbits (MEGNO, \cite{pCi00}), the smaller alignement index and its generalisation (SALI and GALI respectively, see \cite{hSk01,hSk07}), frequency analysis \citep{jLa93}, topological braid chaos methods \citep{jlTh05}, the $0-1$ test \citep{gGo04}, Lagrangian descriptors methods based on lengths of orbits \citep{jDa22,mHi22,jDa23} (requiring the trajectory and the knowledge of nearby trajectories), Birkhoff averages \citep{eSa20}.\\

The remainder of the paper is structured as follows. 
In Sec.\,\ref{sec:methods}, we present the model and computational tools that form the basis of our analysis.
In Sec.\,\ref{sec:DIVvsTime}, we explore numerically the asymptotic time behavior of $DIV$ and reveal distinct power laws of the average $\langle DIV\rangle$ depending on the nature of the orbit, being regular or chaotic.
Our results follow from an extensive parametric investigation based on the standard map model. The power laws we reveal are obtained not only in the original two dimensional phase space but also in reconstructed phase spaces, following a nonlinear time series perspective (\ie by starting from the knowledge of observables). 
In Sec.\,\ref{sec:resultsNOR}, we demonstrate that $DIV$ can effectively be used as chaos indicator. We assess the performances of the $DIV$ measure in the original $2$ dimensional phase space against the fast Lyapunov indicator,  a variational method based on the tangent map.  
Finally, Sec.\,\ref{sec:ccl} summaries and closes the paper. 

\section{Model and computational tools}\label{sec:methods}

\subsection{The standard map}
The standard map is a \tred{two}-dimensional area preserving map paradigmatic of Hamiltonian chaos. The map is obtained as the Poincar\'e map of the periodically kicked rotator model \citep{bCh79}.   
Given a point $(p_{0},\theta_{0})$ on the two dimensional torus $\mathbb{T}=[0,2\pi]^{2}$, 
the dynamics for $n \ge 0$ reads 
\begin{align}\label{eq:StandardMap2D}
	\left\{
	\begin{aligned}
		&p_{n+1}=p_{n}+K \sin \theta_{n} \pmod{2\pi}, \\
		&\theta_{n+1}=\theta_{n}+p_{n+1} \pmod{2\pi}, 
	\end{aligned}
	\right.
\end{align} 
where $\tred{K > 0}$ is the nonlinearity parameter. When $K=0$, the dynamics  is easily described. The ``action'' $p$ is constant and the angle $\theta$ evolves linearly with time. 
When $K > 0$, resonances grow in size and chaos manifests \citep{jMe08}. 

\subsection{Computation of the divergence $DIV$}
\begin{figure}[t]
	\centering
	\includegraphics[width=1\linewidth]{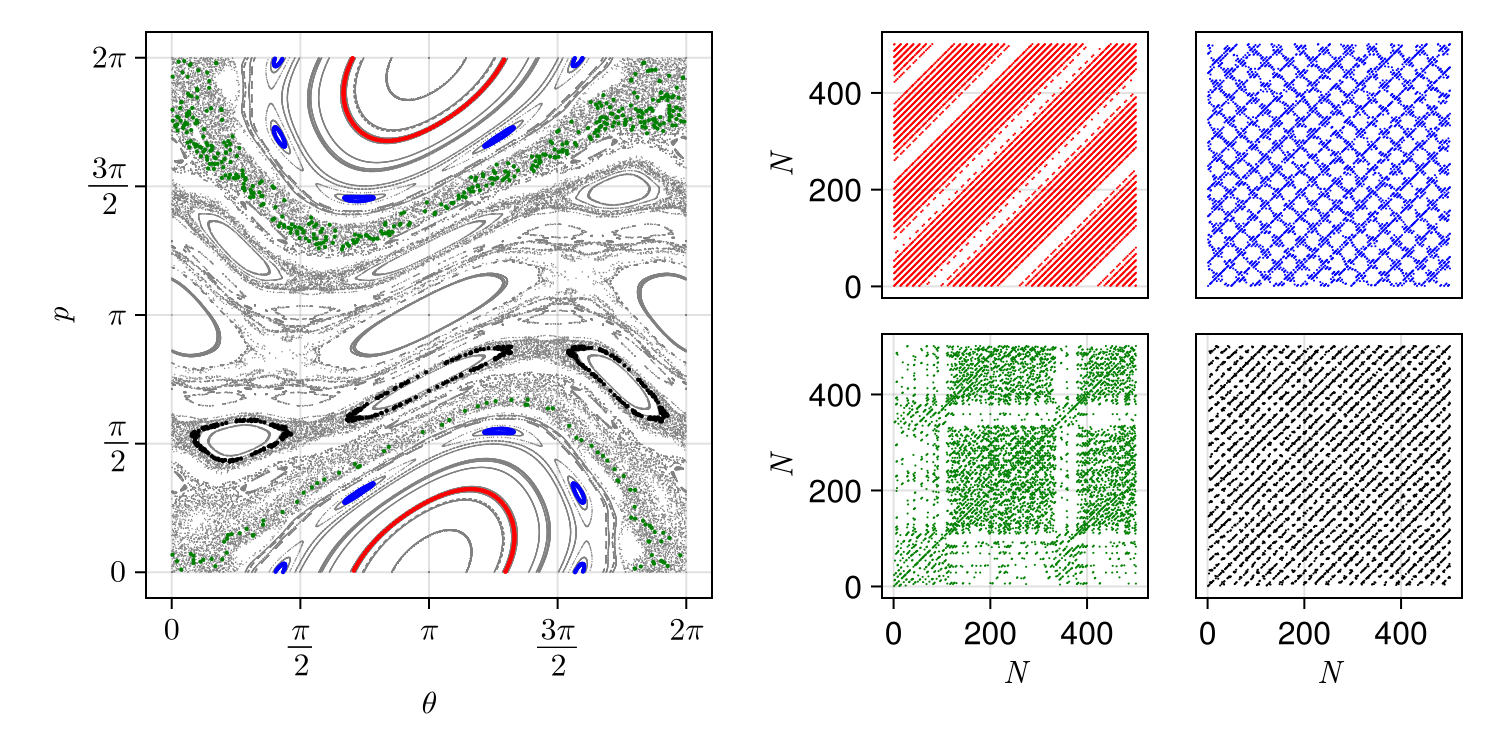}
	\caption{
		(Left) The phase portrait of the standard map at $K=1$ for $N=500$ iterations. 
		(Right) Recurrence plots associated to four representative trajectories shown in color in the phase space; red - quasi-periodic orbit $(\theta_0,p_0)=(3.5,1.0),$ blue - orbit trapped in a secondary resonance $(\theta_0,p_0)=(3.85,1.7),$ green - large scale chaotic orbit $(\theta_0,p_0)=(3.3,1.8),$ black - sticky orbit $(\theta_0,p_0)=(3.14,2.215).$ 
		The different textures in the RPs serve as a basis for recurrence quantification analysis. In particular, we focus here on the divergence  \tred{quantifier} $DIV$, associated to the non-trivial longest diagonal line in a given RP.}
	\label{fig:fig1}
\end{figure}

The divergences $DIVs$ follow from the maximal line length in the diagonal direction of the RPs.  
The RPs and the RQAs measures are computed using the Julia language and specific packages \citep{Julia-2017,Julia-2018}. Several parameters are involved during the computation of the RPs matrices, provided here for the sake of reproducibility. Unless otherwise stated, the norm used in Eq.\,(\ref{eq:rijRP}) is the Euclidean norm. 
To minimise arbitrariness, the parameters \tred{$\epsilon_{i,j}$} follows from a normalised value, a fixed recurrence rate (RR), 
\begin{align}
	RR = \frac{1}{N^{2}}\sum_{i,j} r_{i,j},
\end{align}
here set to $RR=5\%$.   
For RPs computed up to time $N$, the longest diagonal line is the line of identity (trivial diagonal of recurrences). 
This line is excluded from the RPs, together 
\tred{with a Theiler window of size  $w=2$, \ie a band of size $w=2$ around it. Mathematically, this reads 
	\begin{align}
		R_{i,j}=0,  \, \,  i,j=1,\cdots,N,
	\end{align}	
	whenever  $\vert i-j \vert  \le w.$
}	 
The smallest diagonal line \tred{$\ell_{\min}$} considered in RPs is $\ell_{\min}=2$. The Fig.\,\ref{fig:fig1} is an illustrative composite plot showing the phase space of the standard map at $K=1$ together with several RPs for four orbits with distinct properties: an orbit immersed within the main resonant island, an orbit in a secondary resonance, an orbit experiencing stickiness and finally one orbit experiencing large scale chaos.

\subsection{The fast Lyapunov indicator}
\begin{figure}
	\centering
	\includegraphics[width=0.7\linewidth]{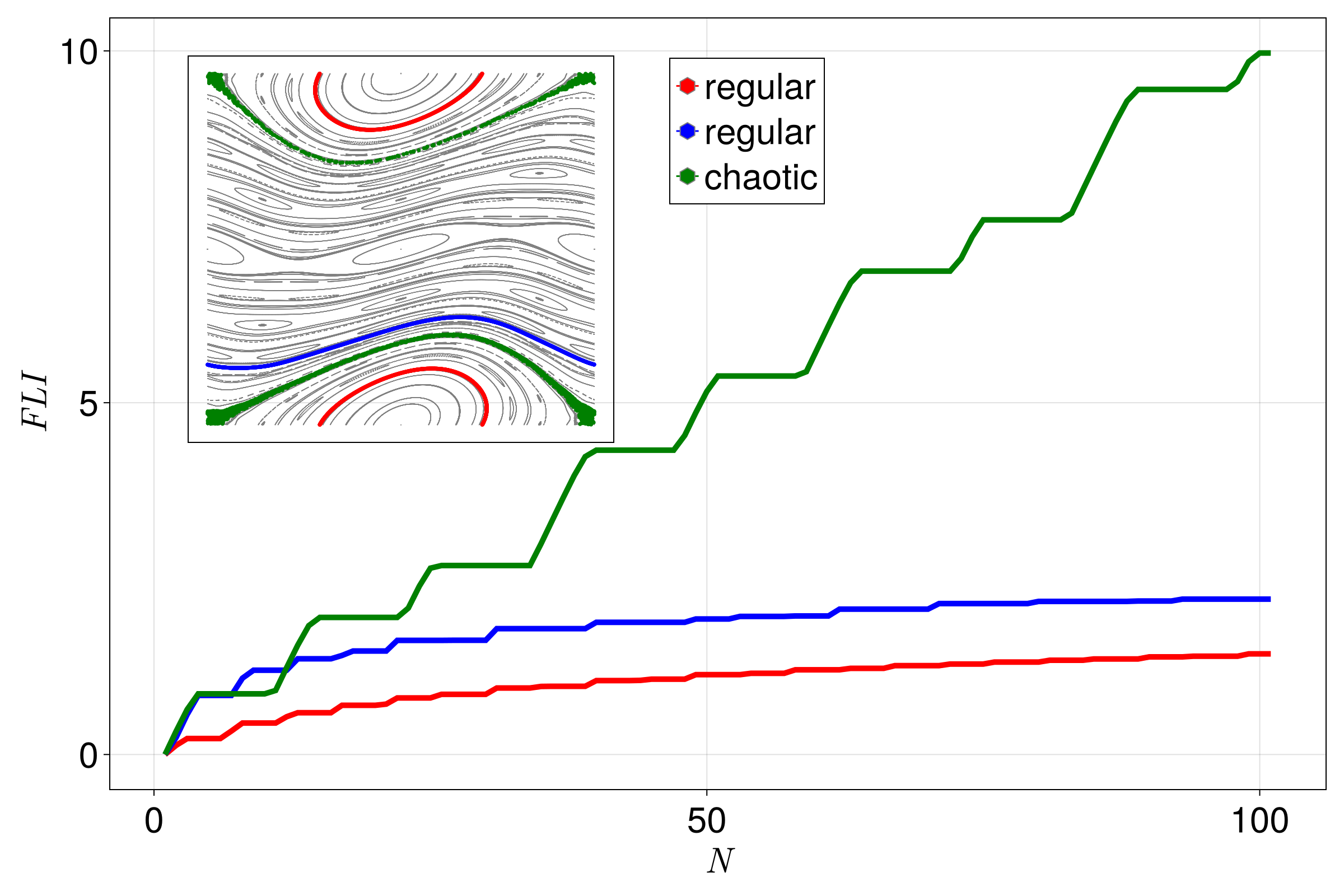}
	\caption{
		\tred{Time evolution of the fast Lyapunov indicator  for a chaotic, circulational and librational  orbit (respectively in green, blue, red) of the standard map with $K=0.6$. 
			The distinct $\mathcal{O}(N)$ and $\mathcal{O}(\log N)$ growth rates
			allow the separation of chaotic and regular orbits in a short time.} 
	}
	\label{fig:fig2}
\end{figure}

A strictly positive largest finite time Lyapunov exponent (FTLE) is usually considered as signature of deterministic chaos (see \cite{hSk09} for a  review), \ie nearby trajectories diverge exponentially fast in average. Here, to assess the presence of chaos,  we instead compute a closely related quantity, namely the fast Lyapunov indicator, hereafter denoted FLI \citep{cFr97}. 
The FLI
is a well established variational chaos indicator, valid for discrete and continuous systems. 
Contrarily to the FTLE, the FLI does not average  the growths of the tangent vector over time, an average that impedes its fast convergence.  
Given 
a smooth mapping $x_{n+1}=M(x_{n})$, $n \in \mathbb{N}$, let 
us denote by $v_{0}$ a unitary deviation vector. 
The tangent map dynamics reads
\begin{align}
	\left\{
	\begin{aligned}
		&x_{n+1}=M(x_{n}), \\
		&v_{n+1}=DM(x_{n})v_{n}.  
	\end{aligned}
	\right.
\end{align} 
There are several definitions of the FLI literature, and in the following we follow the one of \cite{eLe01} that suppresses oscillations,   
\begin{align}
	\textrm{FLI}(x_{0},v_{0};N) = 
	\sup_{n \le N} \log
	\norm{v_{n}}.
\end{align}
For regular orbits, FLI grows as $\mathcal{O}(\log N)$. 
For chaotic orbits, 
the norm of the deviation vector grows exponentially fast, and thus the FLI 
\tred{evolves linearly with time} as $\mathcal{O}(N)$. 
Those two distinct time evolution allow the detection of chaos in a short time \citep{mGu23}. \tred{The time evolution of the FLI is exemplified for regular and chaotic orbits of the standard map in Fig.\,\ref{fig:fig2}}.

\section{Time behavior of $DIV$}\label{sec:DIVvsTime}
Aiming to harness the potential of the $DIV$ measure  as chaos indicator, we first need to ensure that the  \tred{quantifier} reacts  differently on regular and chaotic trajectories. We have investigated time asymptotic properties of $DIV$ on three scenarios, detailed subsequently, by probing parametrically the dynamics of the standard map for many different initial conditions. Our parametric study focuses on properties in the original two-dimensional phase space, but also in reconstructed phase spaces when considering generated time series. 
Quite interestingly, our numerical results highlight clear distinct power laws underpinning the time behavior of the spatial average $\langle DIV \rangle$ for regular or chaotic trajectories.  In the following, the final time $N$ is set to $N=25,000$, considered as here as an ``asymptotic time.''

\subsection{Time behavior of $\langle DIV \rangle$ in the original phase space}\label{subsec:DIVFull}
\begin{figure}
	\centering
	\includegraphics[width=0.8\linewidth]{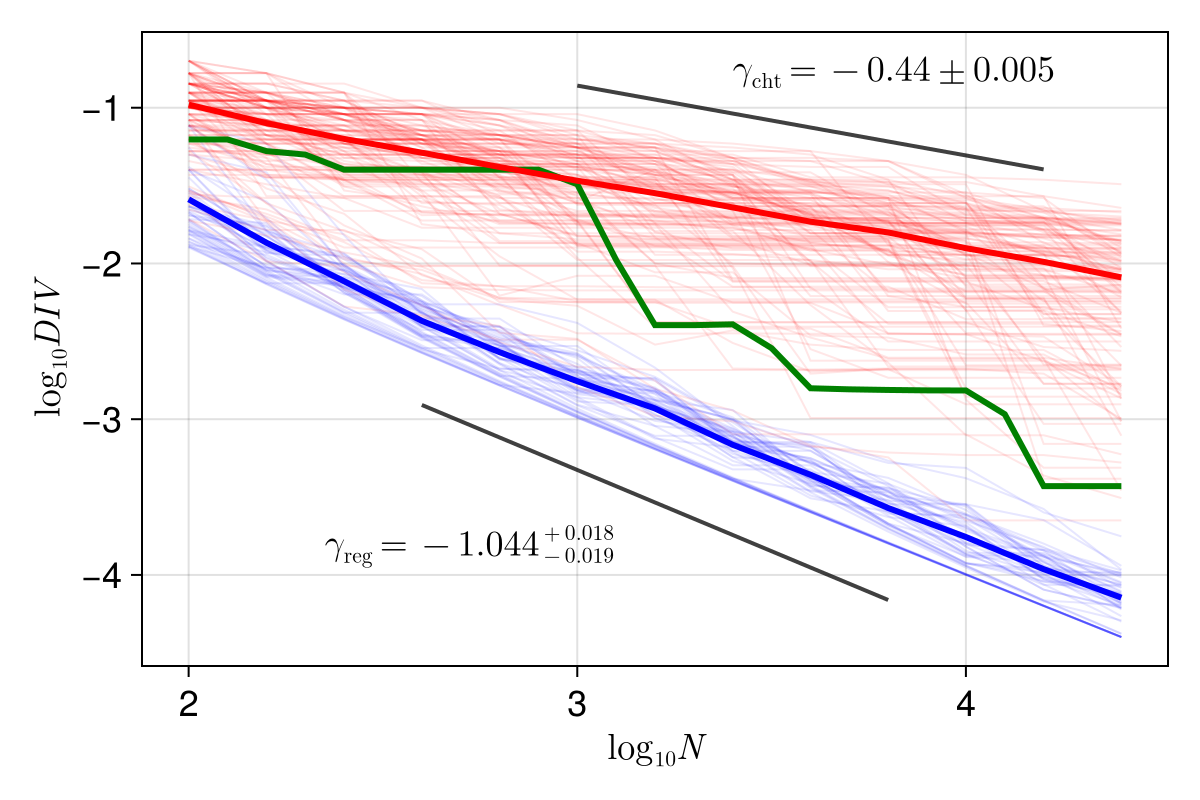}
	\caption{
		\tred{Time evolution of the divergence $DIV$ as a function of the time-series length 
			$N$  for an ensemble of $200$ trajectories of the standard map with varying $K \in [0.6,4]$. Regular orbits are shown in blue and chaotic trajectories in red. The spatial averages $\langle DIV \rangle$
			over the regular and chaotic components (bold blue and red lines, respectively) exhibits clear power law behaviors, whose exponents are estimated numerically. The green curve corresponds to the time evolution of the divergence of a sticky orbit and will be discussed later.}}
	\label{fig:fig3}
\end{figure}

We computed the $DIV$ for a set of $200$ initial conditions distributed randomly into \tred{the domain} $[0,2\pi]^{2}$ for parameters $K$ \tred{also} randomly chosen in the range $[0.6,4]$.  
By using the FLI computed at the final time $N$, each initial condition is  assigned to a specific label (regular or chaotic). The Fig.\,\ref{fig:fig3} shows the time evolution of the divergences $DIVs$ for this set of initial conditions. Initial conditions that are FLI regular are color coded in blue, while  FLI chaotic trajectories appear in red. 
We observe that the $DIVs$ tend to form two distinct clusters, indicating that the $DIV$  \tred{quantifier} effectively captures the regular or chaotic nature of the orbit through its distinct time evolution.
We note the spreading of the $DIVs$ values to be less pronounced in the case of regular orbits. \\

\tred{The spatial averages $\langle DIV \rangle$ of the divergences on the regular and chaotic components  are the central objects of the remainder of this section. They appear a solid bold lines in Fig.\,\ref{fig:fig3}. Interestingly enough they both posses  clear linear trends.}
Their best fits, \tred{determined with a Markov Chain Monte Carlo framework}, lead indeed to power laws  with well  separated exponents \tred{(the posterior medians)}. 
Regular orbits are characterised by 
\begin{align}
	\langle DIV(N) \rangle_{\tred{\textrm{reg.}}} \propto N^{\gamma_\textrm{reg.}}, \, \gamma_{\textrm{reg.}} \sim -1,
\end{align}
numerically, we find $\gamma_{\textrm{reg}}=-1.044$.  Chaotic orbits  experience a much slower decay rate,
\begin{align}
	\langle DIV(N) \rangle_{\tred{\textrm{cht.}}} \propto N^{\gamma_{\textrm{cht.}}}, \, \gamma_{\textrm{cht.}} \sim -1/2,
\end{align}
numerically, we find $\gamma_{\textrm{cht}}=-0.44$. \tred{The 16th and 84th percentiles of the samples  all show a narrow gap. In both cases, the $R^{2}$ coefficients between the space averages and the power laws are $R^{2}=0.99$.}
\tred{We underline} that the asymptotic decay of $\langle DIV \rangle_{\textrm{reg.}}$  echoes the asymptotic decay of the maximal  Lyapunov characteristic exponent \tred{of regular trajectories} \citep{gBe76,gCo78}.   \\

\begin{figure}
	\centering
	\includegraphics[width=0.75\linewidth]{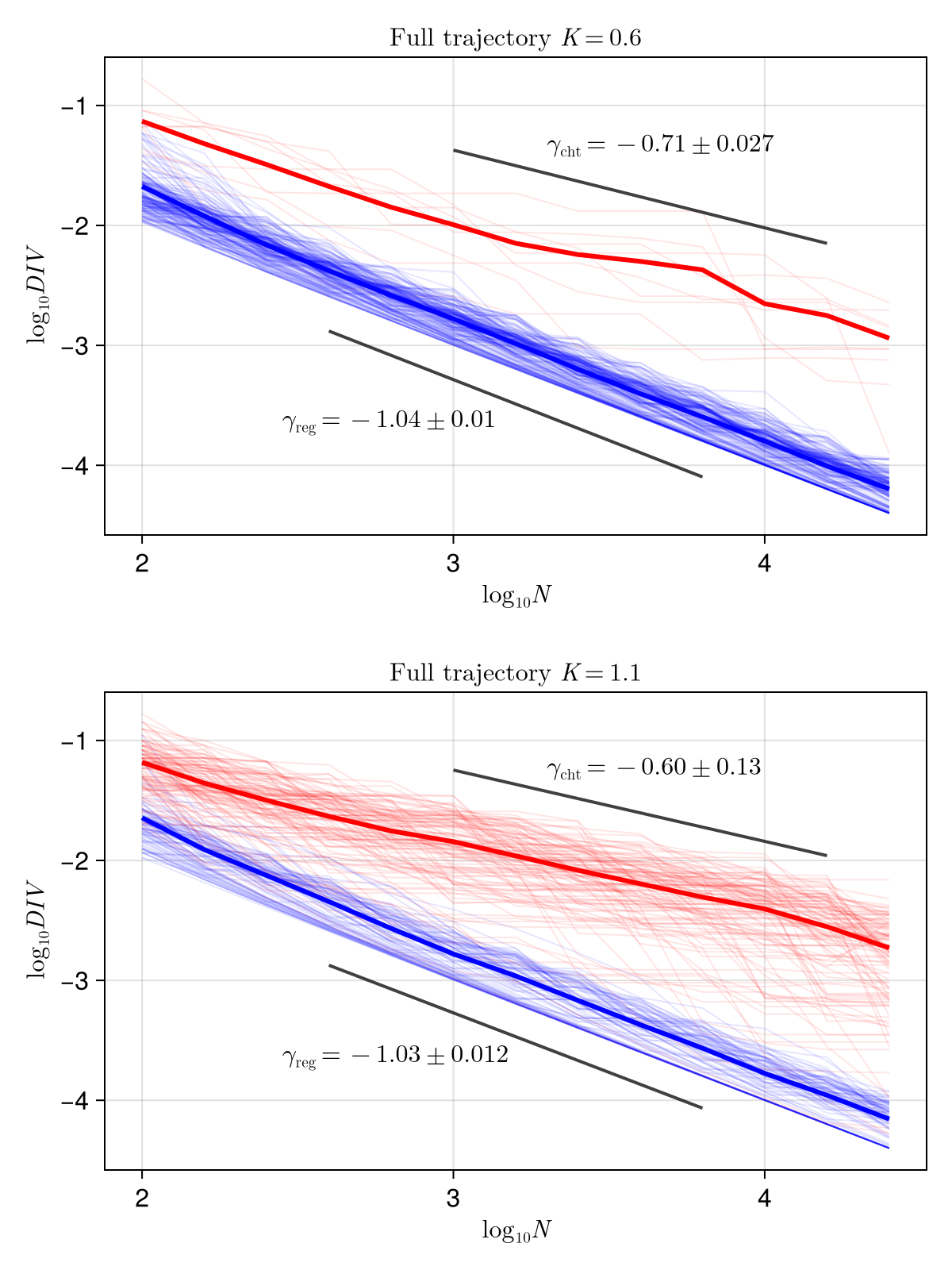}
	\caption{
		Same as in Fig.\,\ref{fig:fig3} for \tred{frozen} parameters $K$. 
	}
	\label{fig:fig4}
\end{figure}

\begin{figure}
	\centering
	\includegraphics[width=0.75\linewidth]{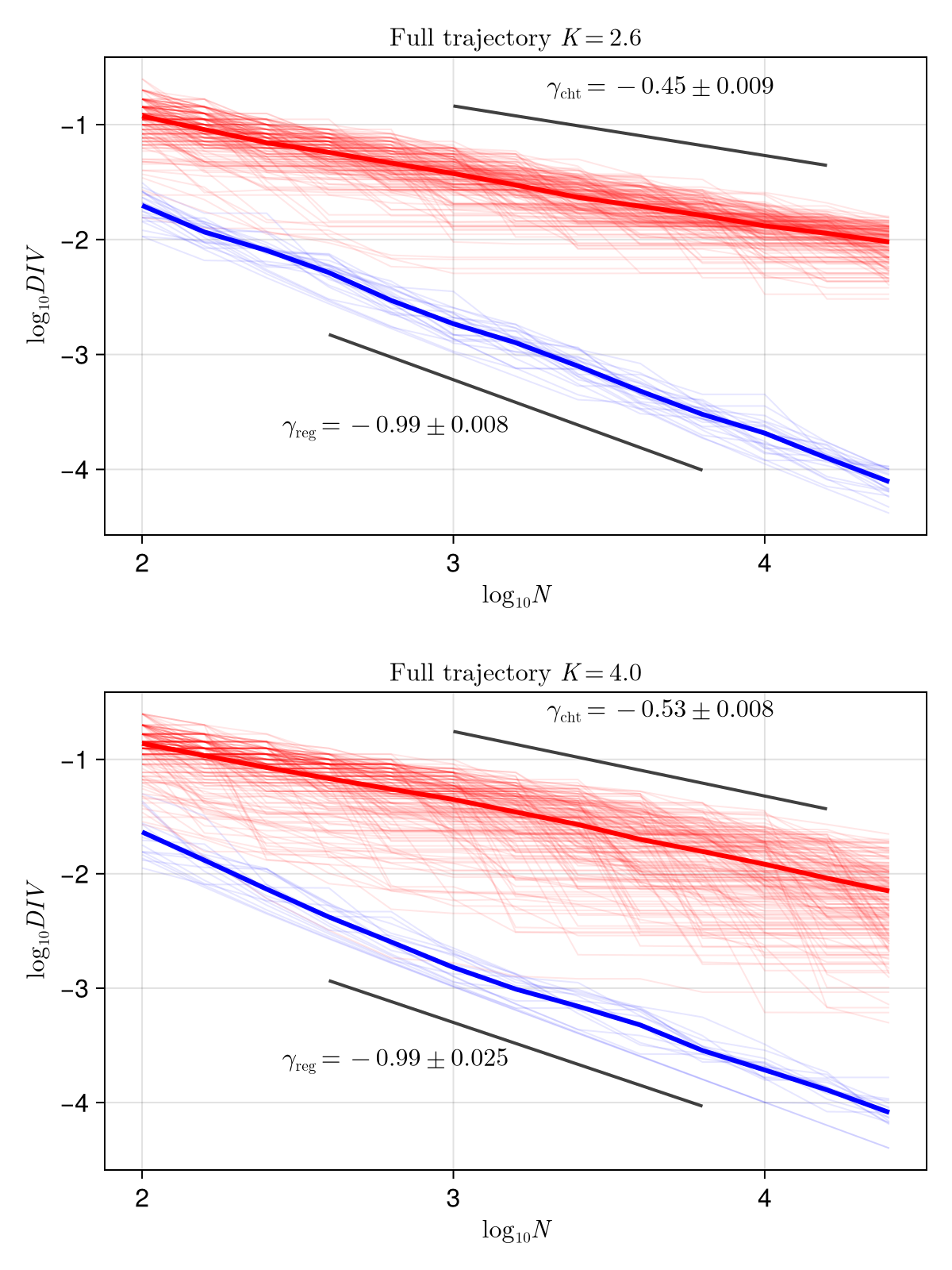}
	\caption{
		Same as in Fig.\,\ref{fig:fig3} for \tred{frozen} parameters $K$. 
	}
	\label{fig:fig5}
\end{figure}

We further confirmed our observations by repeating the previous experiment for $4$ frozen values of $K$, namely $K \in \{0.6,1.1,2.6,4\}$.   
The chosen values are representative of phase spaces dominated by stability ($K=0.6$) or chaoticity ($K=4$). The intermediate value of $K=1.1$ reflects a mixed phase space regime, where chaotic and regular structures cohabit in a balanced proportion.   
Our previous observations fully extrapolate to these $4$ cases. 
In particular, regular trajectories have their $DIVs$ following closely the curve
$N^{\gamma_{\textrm{reg}}}$, with $\gamma_{\textrm{reg}} \le -1$, 
whilst chaotic trajectories have in average a slower decay rate, close to $-1/2$, as shown in Fig.\,\ref{fig:fig4} and Fig.\,\ref{fig:fig5}.  \\

\begin{figure}
	\centering
	\includegraphics[width=0.85\linewidth]{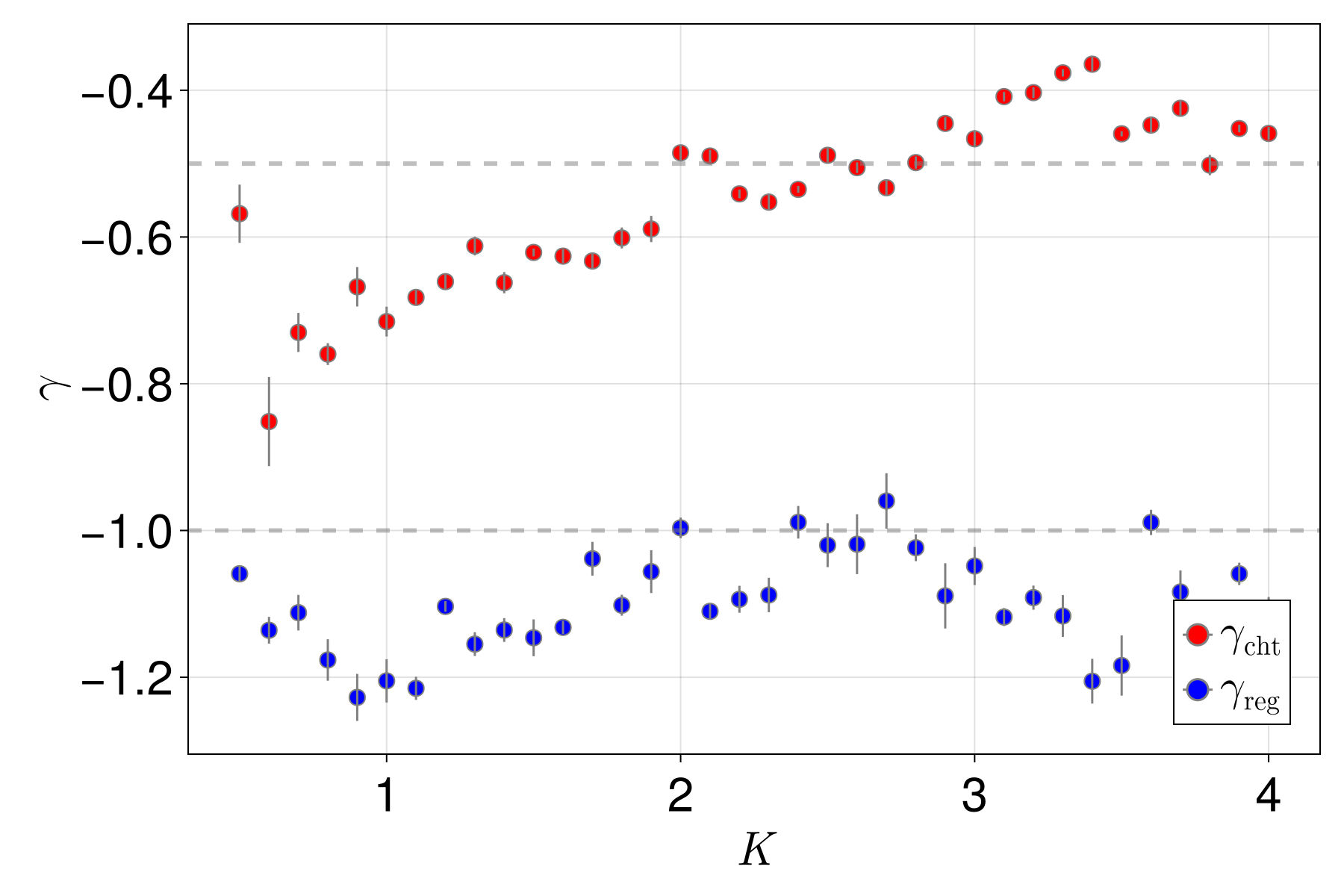}
	\caption{
		Evolution of the power laws exponent $\gamma$ \tred{on the}  regular and chaotic \tred{components} as a function of $K$ computed at $N=500$. The separation between regular and chaotic orbits is well reflected by the \tred{different} exponents, \tred{here shown with their 16th and 84th percentiles (vertical bars)}.  
		For regular components, the exponent is  \tred{most of the time} smaller than $\gamma_{\textrm{reg.}} \le 1$, whilst for
		chaotic components, $\gamma_{\textrm{cht.}}$ is much larger and tends to align to $\gamma_{\textrm{cht.}}=-1/2$. \tred{The two dashed lines, $\gamma=-1$ and $\gamma=-1/2$ are plotted as visual guides.}}
	\label{fig:fig6}
\end{figure}

A finer analysis was performed as a function of the nonlinearity parameter $K$, although this required, \tred{for computational convenience}, reducing the final time to $N=500$. We found $\gamma_{\textrm{cht}} \sim -1/2$ to be characteristic of the decay rate of $\langle DIV \rangle$ in the chaotic regime, see Fig.\,\ref{fig:fig6}. 
(The power laws with pronounced exponents seem to be characteristic of deterministic systems. For a stochastic ARMA process, the exponent is much smaller, reflecting the quasi absence of decay rate of $DIV$, see appendix \ref{sec:app1}). 
\tred{We also underline that the 16th and 84th percentiles, on the chaotic component, tend to be smaller the more chaotic is the system (the larger the $K$). On the regular component, the percentiles are rather stable across the whole range of perturbing values.}
\\

\tred{One may legitimately question the sensitivity of the obtained exponents with respect to the choice of parameters of the RPs. The tables \ref{tab:norm} and \ref{tab:RR} present the exponents we found by changing the norms (from $L_{2}$ to $L_{1}$ and $L_{\infty}$ norms) and the
	recurrence rates (decreasing the nominal value $RR=0.05$ to $RR=0.025$ and increasing it to $RR=0.1$). The close similarity of the exponents derived strengthens confidence in the robustness of the power laws with respect to the choice of RPs parameters.
}

\begin{table}
	\begin{tabular}{l|c|c}
		& $\gamma_{\textrm{reg.}}$ & $\gamma_{\textrm{cht.}}$\\
		\hline
		\hline 
		$L_{2}$-norm & \textbf{-1.04} & \textbf{-0.48} \\
		$L_{1}$-norm & -1.04 & -0.46 \\
		$L_{\infty}$-norm & -1.03 & -0.47 
	\end{tabular}
	\caption{\label{tab:norm}Sensitivity of the exponents with respect to the choice of the norm of Eq.\,(\ref{eq:rijRP}).}
\end{table}

\begin{table}
	\begin{tabular}{l|c|c}
		& $\gamma_{\textrm{reg.}}$ & $\gamma_{\textrm{cht.}}$\\
		\hline
		\hline 
		$RR=0.05$ & \textbf{-1.04} & \textbf{-0.48} \\
		$RR=0.025$ & -1.07 & -0.49 \\
		$RR=0.1$ & -1.00 & -0.47 
	\end{tabular}
	\caption{\label{tab:RR} Sensitivity of the exponents with respect to $\epsilon$ in Eq.\,(\ref{eq:rijRP}) fixed through various recurrence rates $RR$.}
\end{table}

\begin{figure}
	\centering
	\includegraphics[width=1\linewidth]{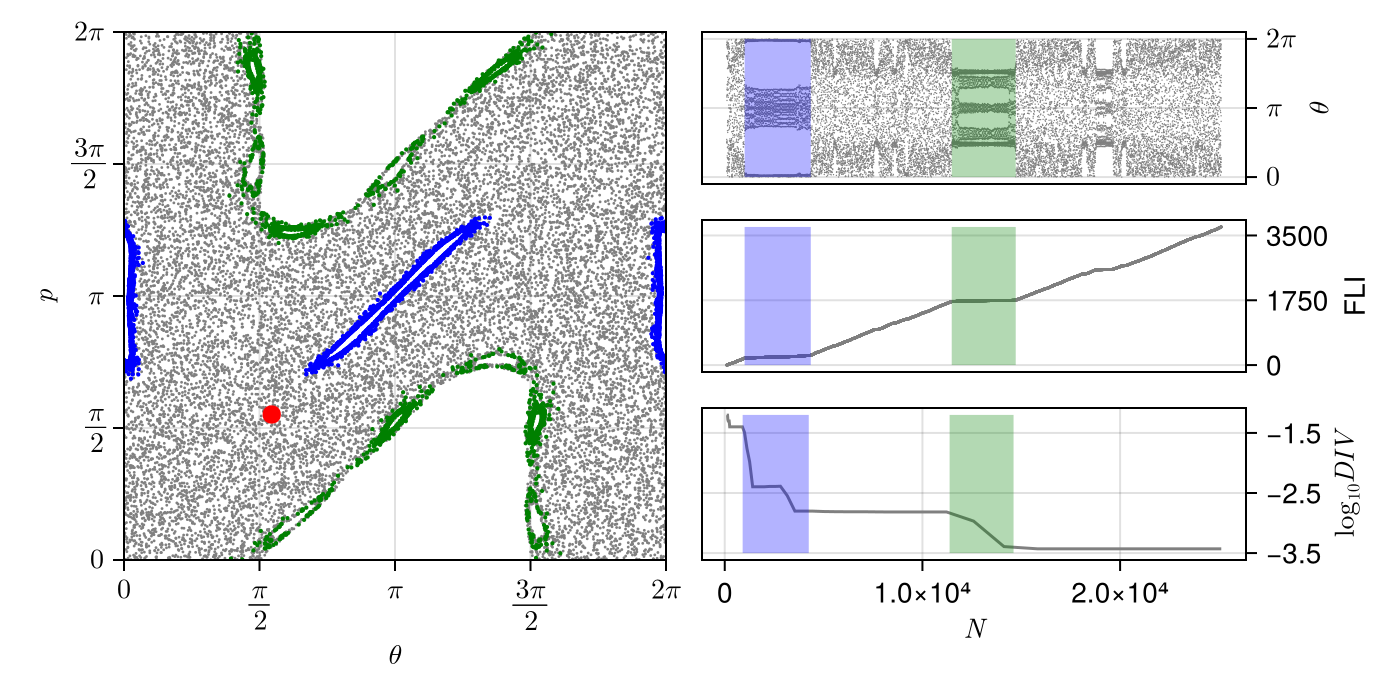}
	\caption{
		(Left)
		Phase space  of the green trajectory of the sticky orbit of Fig.\,\ref{fig:fig3}.  
		Two portions of the orbit are colored (in blue and green) in accordance with the $2$ most prominent sticky events encountered during its evolution. 
		(Right, from top to bottom)
		Time evolution of the angle $\theta$, the FLI and the $DIV$. 
		Sticky events (localised variations of $\theta$) correspond to plateau in the FLI and sharp decay in the $DIV$ measure, as visually materialised with the blue and green boxes serving as visual guides. The sharp decreases of $DIV$ during the sticky events 
		explain the low final value of $DIV$, despite being chaotic, \tred{and the staircase pattern.}
	}
	\label{fig:fig7}
\end{figure}

A close inspection of specific curves  in the set of chaotic trajectories of Fig.\,\ref{fig:fig3} reveals that some orbits have a final $DIV$ comparable to  regular orbits. The green orbit  of Fig.\,\ref{fig:fig3} is one of such orbit. 
We identified initial conditions and parameters of these orbits for a more detailed analysis and found that those orbits exhibit stickiness during their evolution. 
The sharp and successive drops in $DIV$  are correlated to time intervals during which the orbit
is temporarily trapped in specific regions of the phase space. 
During this time, the decay of $DIV$ tends to align to the decay rate of regular trajectories, 
$\gamma_{\textrm{reg.}}=-1$. When the orbit leaves the sticky zone and experiences large excursion, the exponent is closer to $\gamma_{\textrm{cht.}}=-1/2$. Altogether, it produces the apparent staircase pattern. 
This dynamical behavior is also observed and confirmed through the FLI analysis, where the FLI behaves as plateau during during the sticky events (no hyperbolicity contributing to the growth of the norm of the tangent vector). The panel of Fig.\,\ref{fig:fig7} illustrates this phenomenology on a  representative case. 
Thus, we find that the $DIV$ \tred{quantifier} is sufficiently sensitive to detect the presence of sticky dynamics, which manifest as significant fluctuations in $DIV$ over specific time windows corresponding to temporary captures. (Note that this observation provides another RQA measure able to capture stickiness, besides the recurrence rate $RR$, see \cite{mPa22}.)  

\subsection{Time behavior of $\langle DIV \rangle$ starting from an observable, no embedding}\label{subsec:NR}
\begin{figure}
	\centering
	\includegraphics[width=0.75\linewidth]{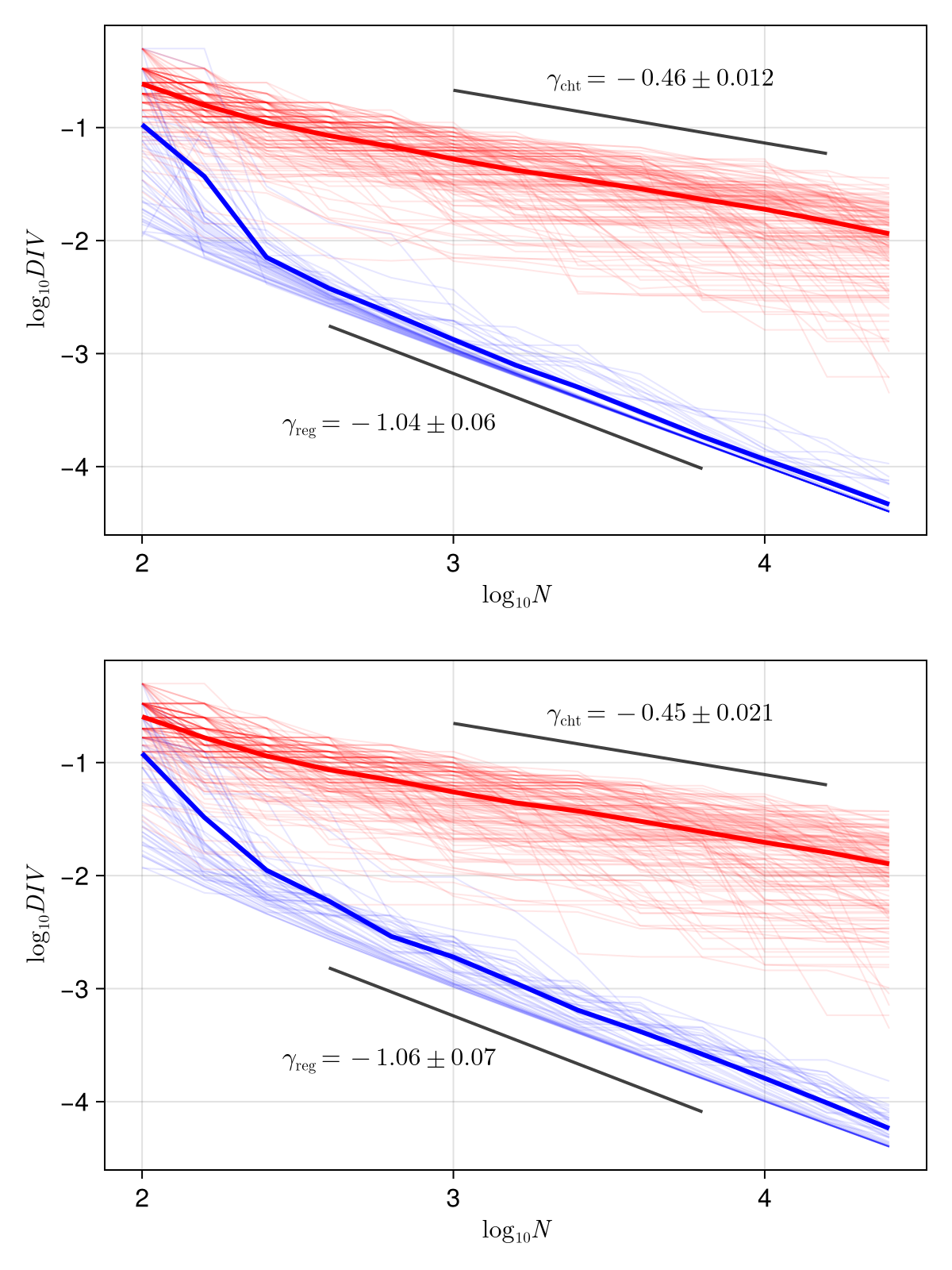}
	\caption{
		\tred{Same as in Fig.\,\ref{fig:fig3}} based on the two observables corresponding to $\theta$
		(top) and momentum $p$ (bottom). 
		No reconstruction of the phase space is performed.  
	}
	\label{fig:fig8}
\end{figure}

The distinct power laws just revealed are based on the computation of the RP from the two-dimensional trajectory. Adopting a nonlinear time series perspective, we now conduct similar numerical experiments starting from the  knowledge of a specific univariate observable $z=g(p,\theta)$ over time, $\{z_{0},z_{1},\cdots,z_{N}\}$, $g$ being an observable function.
The method of delay-coordinates often follows  to reconstruct the phase-space. This is the approach  taken subsection \ref{subsec:delay}. 
For now, we compute the RP plot with no embedding, \ie without any phase space reconstruction.  Given the low-dimensionality of the original system, according to
\cite{jIw98}, it is expected the RP plot to be almost independent from the reconstruction process. As the $DIV$ follows directly from RPs, we thus expect as byproduct the $DIV$ to be also almost independent from the latter.  
Under this setting,  
Eq.\,(\ref{eq:rijRP}) becomes now 
\begin{align}
	r_{i,j}=\Theta\big(\tred{\epsilon_{i}} -\vert z_{i}-z_{j} \vert  \big).
\end{align}
Note that $\norm{\bullet}$ of Eq.\,(\ref{eq:rijRP}) has been replaced by the absolute value.  
We repeated the steps of subsection \ref{subsec:DIVFull} for the following choice of 
observables: $g(p,\theta)=p$ and $g(p,\theta)=\theta$. 
The results are shown in Fig.\,\ref{fig:fig8}. 

\subsection{Time behavior of $\langle DIV \rangle$ starting from an observable, including embedding}\label{subsec:delay} 
\begin{figure}
	\centering
	\includegraphics[width=0.75\linewidth]{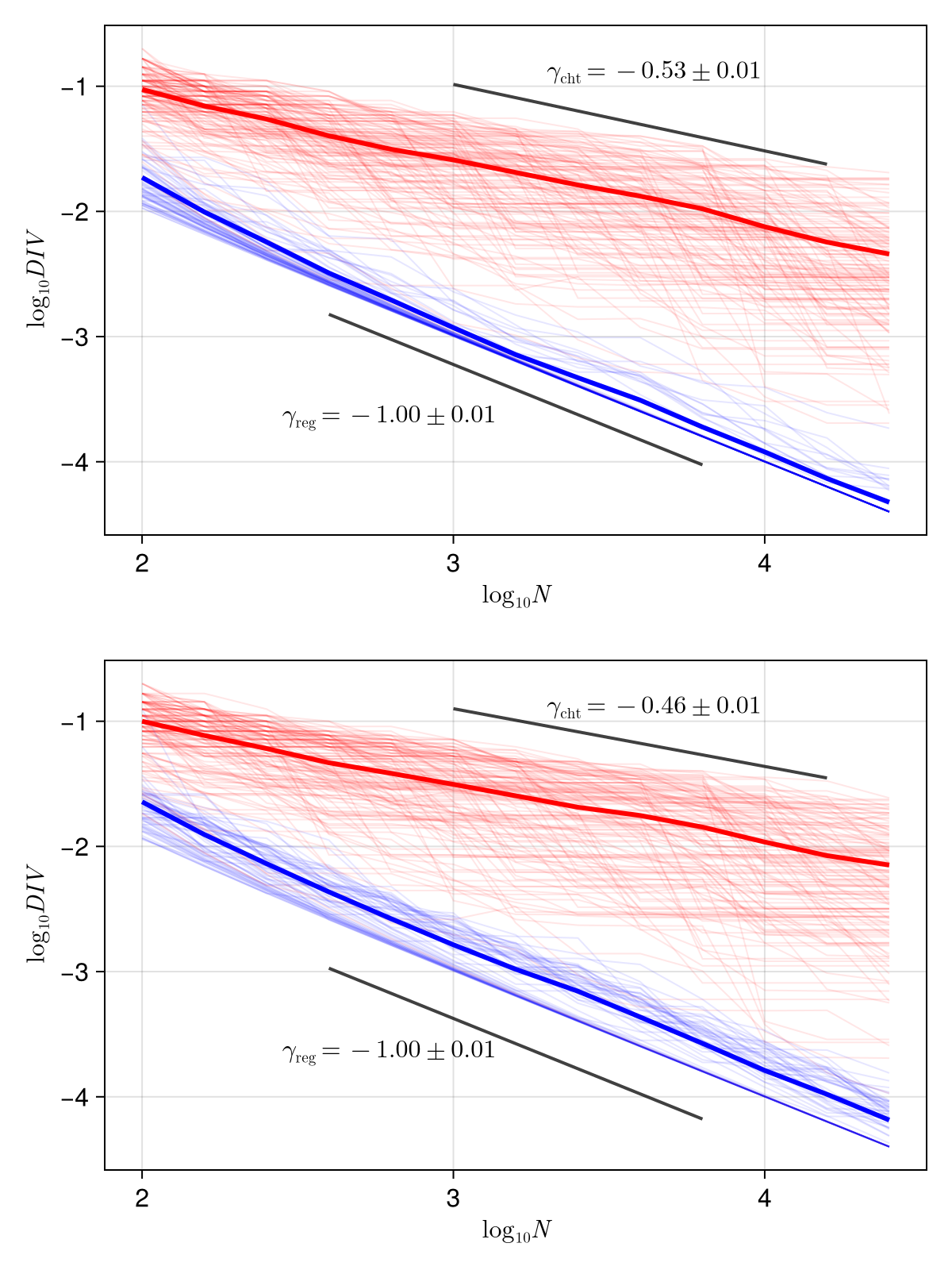}
	\caption{
		\tred{Same as in Fig.\,\ref{fig:fig3}} based on the two observables corresponding to $\theta$
		(top) and momentum $p$ (bottom) \tred{in}
		the reconstructed phase space. 
	}
	\label{fig:fig9}
\end{figure}

Contrarily to the approach taken in subsection \ref{subsec:NR}, starting from the  univariate time series 
$\{z_{0},z_{1},\cdots,z_{N}\}$, 
we now reconstruct a phase space in $\mathbb{R}^{d}$, for some estimated $d$,  
via the delay-coordinates method with lag-time $\tau$ \citep{hAb93}. 
More specifically, let be $m=N-(d-1)\tau$, then 
the reconstructed trajectory reads 
$\{y_{i}\}_{i=0}^{m}$ with
\begin{align}
	y_{i}=(z_{i},z_{i+\tau},\cdots,z_{i+(d-1)\tau}).
\end{align}     
There are several methods to determine the embedding parameters $d$ and delay time $\tau$. 
Here, we fix the time delay to $\tau=1$ and determine $d$ using the false nearest neighbors method \citep{hKa03,mSm05}. 
Once the phase space is reconstructed, 
the divergence is computed from the RP matrix 
\begin{align}
	r_{i,j}=\Theta\big(\tred{\epsilon_{i}}- \norm{y_{i}-y_{j}} \big),
\end{align}
where $\norm{\bullet}$ denotes the Euclidean norm in $\mathbb{R}^{d}$.  
Fig.\,\ref{fig:fig9} presents the analogue of Figs.\,\ref{fig:fig3}, \ref{fig:fig4}, \ref{fig:fig5} and \ref{fig:fig8}  starting from the two observables $z(p,\theta)=\theta$ and $z(p,\theta)=p$ and extrapolates our previous observations.

\tred{
	\subsection{Comments on the generality of the results}
	In subsection \ref{subsec:DIVFull}, we have revealed power laws  with distinct exponents for the divergence quantifier on regular and chaotic components. We presented numerical evidences that our results do not chiefly depend on the parameters of the RPs (norms and recurrence rates). 
	In subsections \ref{subsec:NR} and \ref{subsec:delay}, we extended our conclusions starting from observables, and found similar results in reconstructed phase spaces. 
	All these results are based on the standard map, a mapping derived from a  Poincar\'e map of a continuous time dependent system. In Appendix \ref{sec:app2}, we have considered  Poincar\'e  sections of two archetypal  non-integrable  resonant Hamiltonian models where resonances overlap, with one degree-of-freedom and non-autonomous.  Our results and observations extrapolate to those models too, bringing further confidence to the generality of the observations made on the standard map model.  
}

\section{Performance assessment of $DIV$ as chaos indicator in the original phase space}\label{sec:resultsNOR}
We now qualitatively and quantitatively assess the sensitivity of the $DIV$ measure as finite time chaos indicator. Given the similarities of the power laws revealed in Sec.\,\ref{sec:DIVvsTime} in the original or reconstructed phase spaces, we  focus solely on the performances in 
the original phase space, 
\ie using the  2 dimensional original trajectory.
We benchmark the performance against the FLI, considered here as ``ground-truth.''  
Our assessment is performed at $N=500$, thus working with data lengths that are relevant to real-world scenarios.  \\

Given an initial condition $x_{0}$ and a final time $N$, it is first desirable to define a threshold $\alpha$ to establish a binary classification of the orbit as regular or chaotic from the value $DIV(x_{0};N)$. More precisely, we are looking to a criteria 
\begin{align}\label{eq:ThresholdsDelta}
	\left\{
	\begin{aligned}
		& DIV(x_{0};N) \le
		\alpha \Rightarrow \textrm{``The orbit stemming from $x_{0}$ is regular,''}\\
		& DIV(x_{0};N) >
		\alpha \Rightarrow \textrm{``The orbit stemming from $x_{0}$ is chaotic.''}
	\end{aligned} 
	\right.
\end{align}
To determine the threshold $\alpha$, we follow here  standard approaches used for variational methods relying on the shape of the histogram of the \tred{quantifier} computed for many initial conditions. \tred{(This strategy is inspired   by \cite{jSz05} in the context of finite-time Lyapunov exponent, and also used more recently in \cite{mrS22,mHi22,aCa25} with the Birkhoff average method or Lagrangian descriptors}.)  \tred{This is also the methodology we follow to binarise the FLI outputs.\\}

\begin{figure}
	\centering
	\includegraphics[width=1\linewidth]{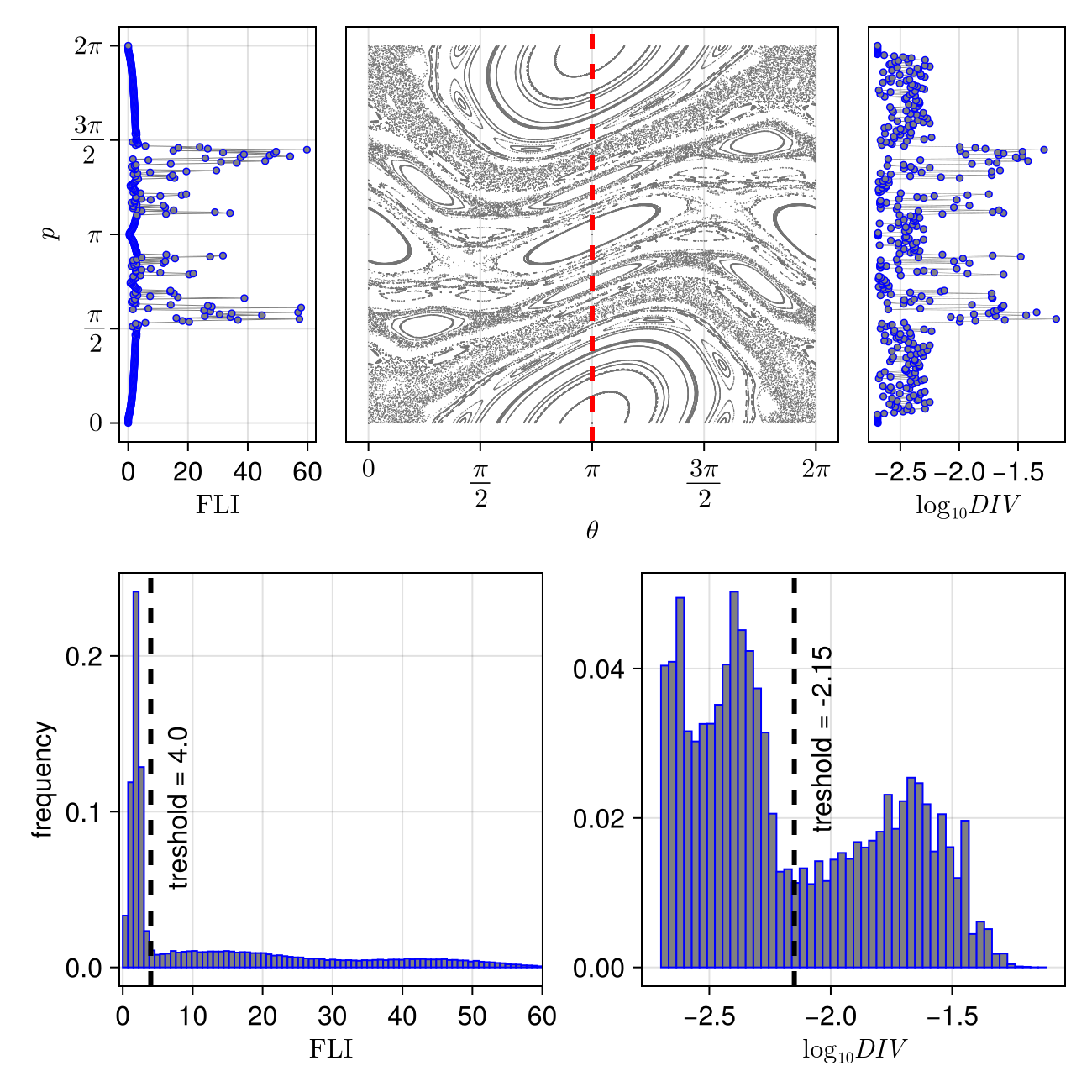}
	\caption{ (Top row) 
		Phase portrait of the standard map at $K=1$ along with  FLI and  $DIV$ landscapes computed at $N=500$ along the dashed vertical red line of initial conditions. 
		(Bottom row)
		Distributions of the FLI and $DIV$
		for $400\times 400$ uniformly distributed initial conditions in $[0,2\pi]^{2}$. 
		Inspection of the histograms allow to set a threshold $\alpha$ to binarily classify orbits as regular or chaotic. See text for details.}
	\label{fig:fig10}
\end{figure}

Fig.\,\ref{fig:fig10} shows the phase space of the standard map for $K=1$  together with the 
landscapes of the FLI and $DIV$ for $250$ initial conditions spread along the dashed line joining the initial conditions $(\theta,p)=(\pi,0)$ and $(\pi,2\pi)$. 
For both indicators, we identify sharp increases, in $1-1$ correspondence, when they cross transversely the thin chaotic layers. 
The bottom row of Fig.\,\ref{fig:fig10} shows the histograms of their final values for a resolved $400 \times 400$ Cartesian mesh of initial conditions in $[0,2\pi]^{2}$. 
The FLI distribution is rather right skewed, with the main peak located close to $\log(N)=\log(500) \sim 2.69$. This value characterizes the background of regular orbits. In this case, setting the threshold value $\alpha$ larger than this furnishes a reliable threshold to separate chaotic from regular trajectories, here set to $\alpha=4$. The distribution of the $DIVs$ is rather bimodal, where each mode reflects the typical values taken on regular and chaotic components.  
In this case, the threshold $\alpha$ is set as the minimal value between the two modes, leading here  to $\alpha=2.15$. 
In the following, we systematically rely on histogram inspection to determine ad hoc $\alpha$ values (for regimes dominated by regularity, the histogram is skew right, for a regime dominated by chaoticity, it becomes skew left). \\

\begin{figure}
	\centering
	\includegraphics[width=1\linewidth]{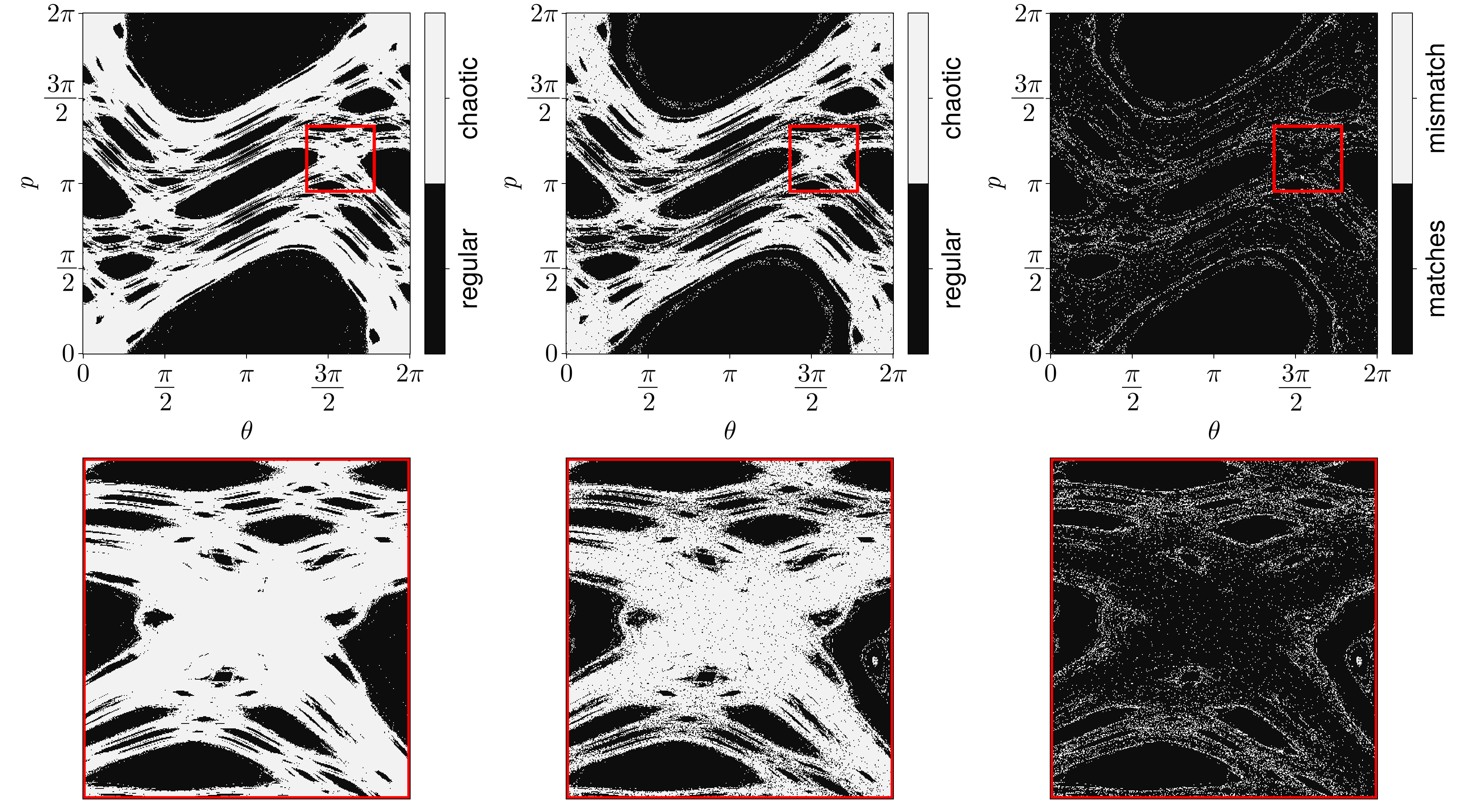}
	\caption{(Top row)
		Stability maps 
		encoding the regularity (black) or chaoticity (white) obtained with the FLI and the $DIV$, respectively. The last figure shows the mismatch points, \ie the points for which the classification disagree (white). (Bottom row)  
		Same analysis at a smaller scale  delineated by the red box in the top row. 
		Most of the mismatch points are located in the vicinity of the edges of the dynamical structures. 
	}
	\label{fig:fig11}
\end{figure}

Fig.\,\ref{fig:fig11} portrays a series of dynamical maps corresponding to a scan of $[0,2\pi]^{2}$ (top row) and a magnification of it (bottom row) materialised by the red square appearing near the $2:1$ periodic orbit. Each domain is meshed with a regular $400 \times 400$ grid of initial conditions. Initial conditions leading to chaotic motions (either with the FLI or $DIV$) are marked in white, whilst regular orbits appear in black. 
Qualitatively, \tred{we found} the dynamical maps of the first two columns of Fig.\,\ref{fig:fig11} \tred{to be very similar}. 
\tred{These results support the validity of the $DIV$ quantifier as an indicator of chaos.} 
By plotting the 
initial conditions for which the labels are in disagreement, we are able to reveal the mismatch set (third column of Fig.\,\ref{fig:fig11}). 
The two-scale analysis of the mismatch sets highlights that mismatch points are predominantly concentrated along the edges of the dynamical structures. \\

\begin{figure}
	\centering
	\includegraphics[width=1\linewidth]{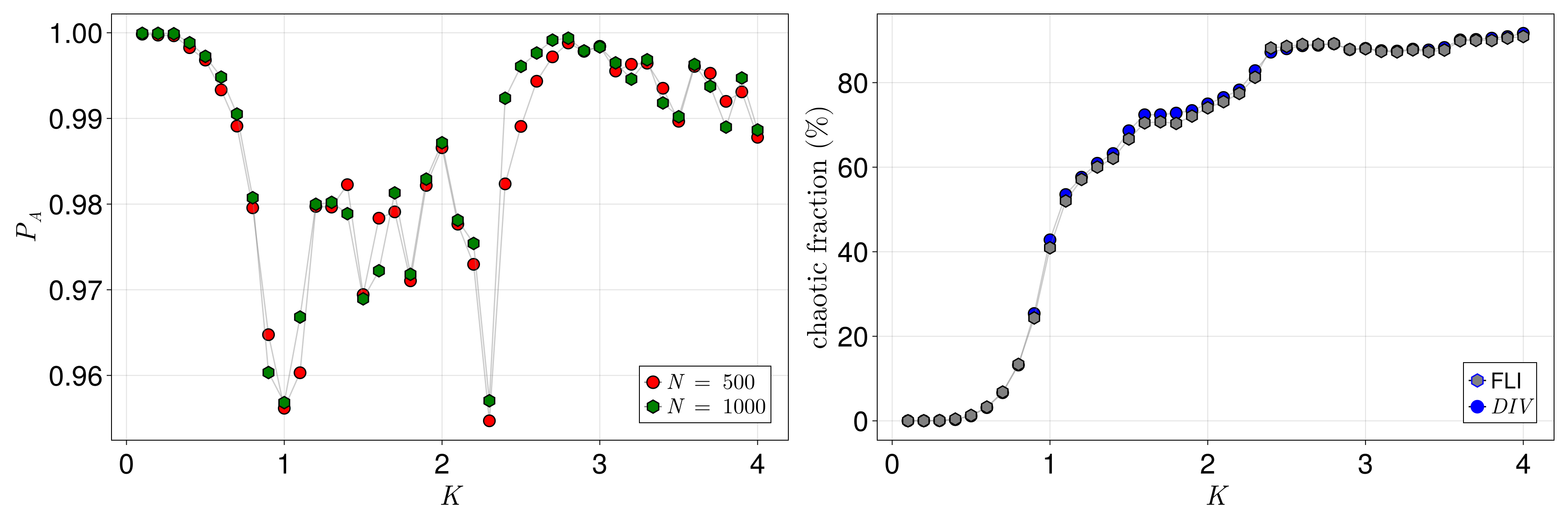}
	\caption{
		(Left) Evolution of the probability of agreement $P_{A}$ (confer Eq.~(\ref{eq:PA})) versus the nonlinearity parameter $K$. $P_{A}$ is computed over the domain  $A=[0,2 \pi]^{2}$. (Right) The size of the chaotic region in the phase space for different $K$ computed either with the FLI or the $DIV$ \tred{on the same domain $A$}. 
		The two curves closely follow each other, strengthening further the use of $DIV$ as chaos indicator. 
		Both panels are based on a mesh of $400\times 400$ initial conditions computed at $N=500.$}
	\label{fig:fig12}
\end{figure}

In order to compare the dynamical maps more quantitatively, we compute a simple metric with a straightforward interpretation: the probability of agreement $P_{A}$ between the labels produced (chaotic or regular) with either method over a specific domain $A$ (discretised as a regular Cartesian mesh of initial conditions.) 
The metric reads
\begin{align}
	P_{A}=
	\frac{\# \textrm{Labels in agreement}}{\# 
		\textrm{Initial conditions in $A$}}.
	\label{eq:PA}
\end{align} 
Note that \tred{the scalar} $P_{A}$ concatenates the outputs of resolved dynamical maps \tred{computed over the domain $A$}. 
For $N=500$, we found the $DIV$ \tred{quantifier} to perform overall very well against the FLI, with \tred{a probability of } agreement over $95\%$ over the range of nonlinearity parameters $[0.6,4]$, as illustrated in the left panel of Fig.\,\ref{fig:fig12}. \tred{For $K <1$, the probability of agreement is over $98\%$, just as for the values of $K>2.5$}.

\tred{For} \tred{the} values of $K \in [1,2.2]$,  \tred{a slight decrease in the agreement probabilities between the indicators is observed. Nevertheless, the agreements are still above $95\%$.} 
These dynamical regimes are characterized by regular and chaotic  structures that cohabit in a rather balanced manner (mixed phase space regime).  \tred{The presence of numerous thin secondary structures  naturally challenges the performance of the indicator, and possibly explain the dip.}
\tred{Extending the final time to 
	$N=1,000$
	does not markedly affect this picture, though it yields a slight improvement in the agreement.}
\tred{To further complement the results,} the right panel of Fig.\,\ref{fig:fig10}  shows, as a function of the nonlinearity parameter $K$, the volume of chaotic trajectories estimated either with the FLI or the $DIV$. 
The computation closely follows the  \tred{steps} of  \cite{eSa20} \tred{with the weighted Birkhoff average}. 
Throughout the range of  \tred{nonlinear parameters},  we observe that the two curves closely follow one another, further demonstrating the effectiveness of the $DIV$ method in serving as a reliable chaos indicator.

\section{Conclusions}\label{sec:ccl}
Although RPs and RQAs have been predominantly applied to dissipative systems, our study demonstrates the effectiveness and potential of the $DIV$ \tred{quantifier} to serve as a simple diagnostic  to detect chaos in conservative systems.  
We have compared the performance of this indicator against the fast Lyapunov indicator, a well established variational chaos detection method. We focused on the standard map, a paradigmatic discrete model of Hamiltonian chaos.
From our extensive numerical simulations, we have quantitatively assessed the overall good agreement between the methods, further strengthening, but this time in the conservative regime, the strong correlation between the divergence measure and the presence of chaos in the dynamics. 
Our comparison has focused on moderately long orbits with $500$ data points, stepping towards real-world applicability of the method.  
The distinct power laws of $DIV$ we have revealed in average depending on the regularity or chaoticity of the trajectories, on a much longer timescale, valid in the original and reconstructed phase space, shed also more light on asymptotic properties of the $DIV$  \tred{quantifier}. 
In particular, we have observed a decay of $DIV$ as $1/N$ for regular orbits, which is interestingly the same rate as the maximal Lyapunov characteristic exponent.  
\tred{On chaotic components, a decay rate closer to $1/\sqrt{N}$ is usually observed.}
These distinct power laws could also be particularly valuable when dealing with a limited number of orbits, a context in which setting the threshold $\alpha$ from the histogram becomes challenging. Those properties suggest a possible new approach for chaos detection in time series, based on analyzing the slope of the decay rate of $DIV$ computed over windowed segments of the original series. 
Our current efforts are focused on leveraging this property, besides the noise free case, alongside providing analytical insights to support our findings.

\section*{Acknowledgments}
This work was supported by the Hungarian National Research, Development and Innovation Office, under Grant No. K-153324, K-152888, and TKP2021-NKTA-64 (T.K.), financed by the Ministry of Culture and Innovation of Hungary.

\section*{Data Availability Statement}
The data presented in this study can be reproduced based on the computational notebooks availabe on Zenodo: \url{https://doi.org/10.5281/zenodo.18356624} \citep{tKo26}. 

\newpage
\appendix 

	\section{$DIV$ versus $N$ for ARMA process}\label{sec:app1}
\begin{figure}
		\centering
		\includegraphics[width=0.7\linewidth]{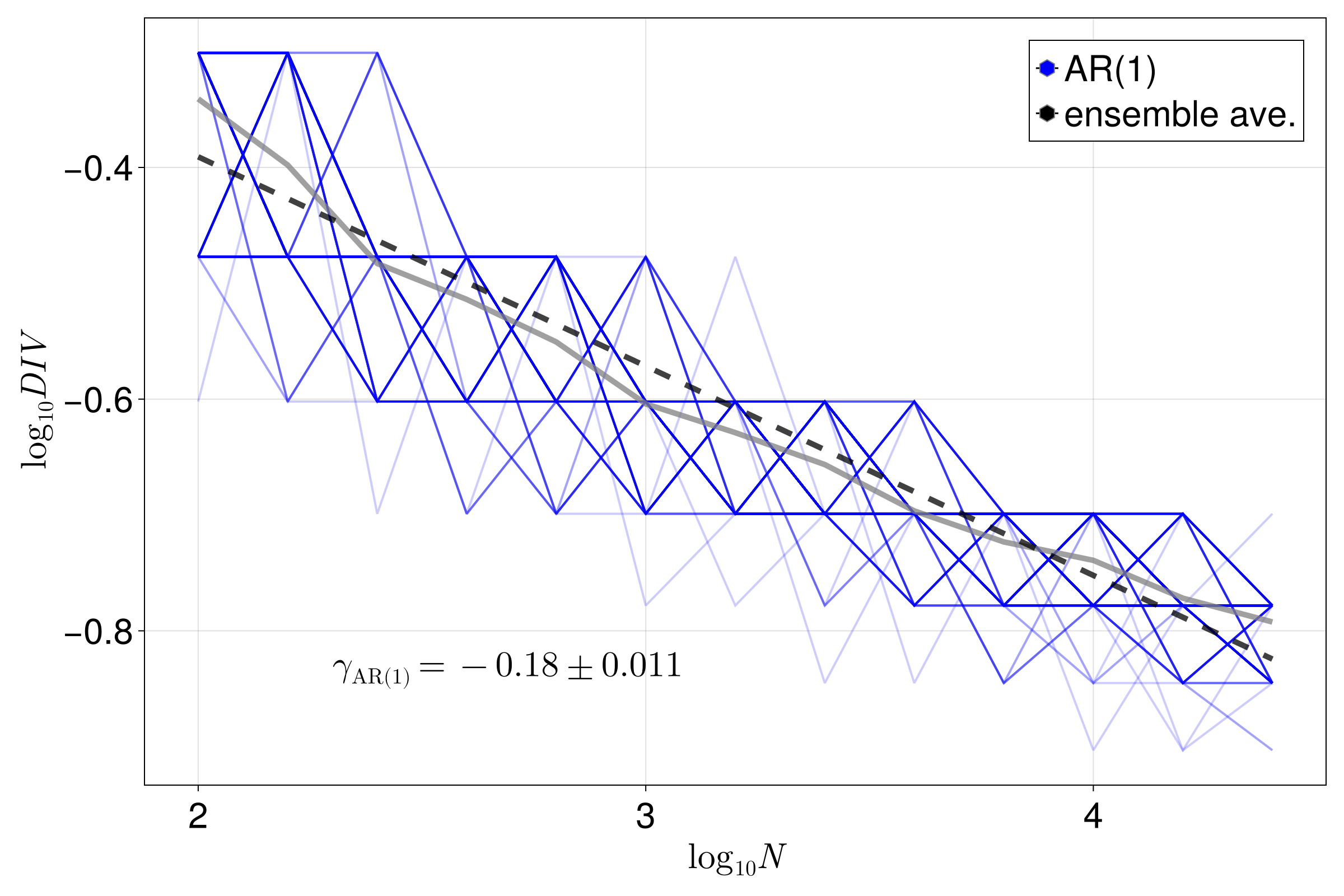}
		\caption{
			Time evolution of $DIV$ for a $\textrm{AR}(1)$ process together with the fit of it's time evolution.  	
		}
		\label{fig:figAp1}
	\end{figure}

	Figure \ref{fig:figAp1} reports the time evolution of an autoregressive $\textrm{AR}(1)$ model. 
	We observe a quasi absence of decay of $DIV$ versus the time. Note that the exponent found is one order of magnitude smaller than characteristic exponents found in the deterministic case. 
	
	\section{Applications to two stroboscopic maps computed from resonant Hamiltonians}\label{sec:app2}
	We report the temporal laws of $DIV$ obtained for two stroboscopic maps of two Hamiltonian models. Both models are time-periodic $1$ degree-of-freedom (DoF) models that contain resonant harmonics. 
	In both cases, the stroboscopic map $P$ is $2$ dimensional and has a cylindrical topology, \ie its variables $\bold{x}=(x,y) \in \Sigma = S^{1} \times I$, $I \subset \mathbb{R}$ and $S^{1}$ is the circle. 
	Let be 
	\begin{align}
		\Sigma_{T} = \{(\bold{x},t) \in \Sigma \times \mathbb{T}_{T}\}, \, \mathbb{T}_{T}=\mathbb{R}/T\mathbb{Z}.
	\end{align}
	For an initial condition $z=(\bold{x},t) \in \Sigma_{T}$, we consider the $T$-time map 
	\begin{align}
		\Phi^{T}:  & \, \Sigma_{T} \to \Sigma_{T}  \\ \notag
		&  \,  z \mapsto  z'=(\bold{x}',t')=\Phi^{T}(z).
	\end{align}
	The stroboscopic map then reads
	\begin{align}
		P:  & \, \Sigma\to \Sigma  \\ \notag
		&  \,  \bold{x} \mapsto  P(\bold{x})=\bold{x}'.
	\end{align}
	Presentation of the results follow closely Sec.\,\ref{subsec:DIVFull}. In particular, the computation of $DIV$  is performed in the original 2-dimensional phase space of the map $P$ (no reconstruction whatsoever). 

	\subsection{Results for two resonances that overlap}
	We consider the model 
	\begin{align}
		\mathcal{K}(I,\phi,t)
		=
		\frac{I^{2}}{2}-
		(\alpha_{1}\cos(\phi-t) 
		+\alpha_{2}\cos(\phi+t)) \label{eq:modK}
	\end{align}
	\tred{discussed in \cite{aMo22} (chapter $6$).}
	For this model, $T=2\pi$. 
	\tred{
		The 2-DoF autonomous counterpart of Eq.\,(\ref{eq:modK}) reads
		\begin{equation}
			\mathcal{K}
			=
			\frac{I^{2}}{2}+J-
			(
			\alpha_{1}
			\cos(\phi-\tau)+
			\alpha_{2}\cos(\phi+\tau)
			).
		\end{equation}
		When $\alpha_{1}=0$ or $\alpha_{2}=0$, we recover the unperturbed integrable Hamiltonian of the pendulum using an ad-hoc canonical change of variables.  
		The phase space contains the single cat-eye resonance centered around either $c_{1}(I)=-1$ or $c_{2}(I)=1$, with half-widths of the separatrices $\delta_{2}=2\sqrt{\alpha_{2}}$ or $\delta_{1}=2\sqrt{\alpha_{1}}$ respectively. 
		Whenever both $\alpha_{1}$ and $\alpha_{2}$ are different from zero, the resonances might overlap and chaos appears.  
		The resonance overlap  parameter, also called \textit{stochasticity parameter} \citep{bCh79}, reads
		\begin{equation}
			s=\frac{\delta_{1}+\delta_{2}}{\vert c_{2} - c_{1}\vert}=\sqrt{\alpha_1} + \sqrt{\alpha_2}.
		\end{equation}
		In our numerical setting, we assign to each resonant eye the same dynamical weight with $\alpha_{1}=\alpha_{2}=1/5$, leading to $s\approx 0.89$.  As $s$ is close to $1$, the resonances overlap significantly and macroscopic chaos is expected. 
	}
	
\begin{figure}
		\centering
		\includegraphics[width=1.\linewidth]{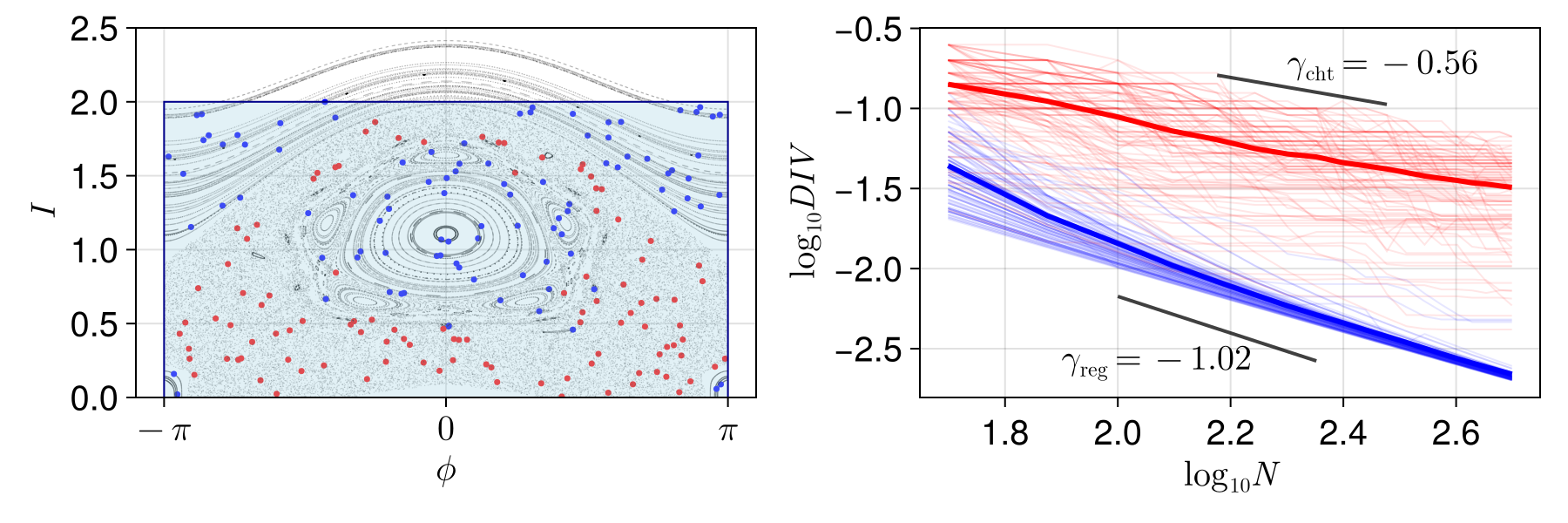}
		\caption{
			(Left) Phase portrait for the resonance overlap Hamiltonian (\ref{eq:modK}) \tred{obtained with its associated Poincar\'e map}. (Right) \tred{The spatial averages $<DIV>$ follow distinct power laws on the regular and chaotic components.}}
		\label{fig:figA2}
	\end{figure}

	The phase space \tred{obtained by iterating} $P$ together with the temporal laws of $DIV$ on regular and chaotic components are shown in Fig.\,\ref{fig:figA2}.  \tred{The fits of the spatial averages lead to} $\gamma_{\textrm{reg}.}=-1.02$ and $\gamma_{\textrm{cht}.}=-0.56$ \tred{in the original two-dimensional phase space of $P$}.\\
	
	\tred{Although we do not show the various plots here, Tab.\, \ref{tab:modK} summarises the exponents we found under various dynamical setting as we did in Sec.\,\ref{sec:DIVvsTime}, \ie by computing the $DIVs$ from observables, including or not the phase-space reconstruction. The exponents found are in accordance with the exponents determined so far.}
	
	\begin{table}
		\begin{tabular}{l|c|c}
			& $\gamma_{\textrm{reg.}}$ & $\gamma_{\textrm{cht.}}$\\
			\hline
			\hline 
			Original $2$D phase space & -1.02 & -0.52\\
			Observable $I$, reconstruction & -1.06 & -0.62 \\
			Observable $\phi$, reconstruction & -1.03 & -0.62\\
			Observable $I$, no reconstruction& -1.16& -0.59\\
			Observable $\phi$, no reconstruction & -1.06 & -0.59
		\end{tabular}
		\caption{
			\label{tab:modK}
			Exponents for the power laws of $\langle DIV \rangle$ found for the Poincar\'e map associated to the continuous model $\mathcal{K}$ of Eq.\,(\ref{eq:modK}) under different dynamical settings. The denomination ``observable $I$'' and ``observable $\phi$'' is an abuse of notation to refer to the discrete variables of the Poincar\'e mapping associated to the original momentum $I$ and angle $\phi$. The parameters according to time-delay phase space reconstruction \citep{fTa81} are $d=5$ and $\tau=1.$
		}
	\end{table}
	
	\subsection{Results for a modulated pendulum}
\begin{figure}
		\centering
		\includegraphics[width=1\linewidth]{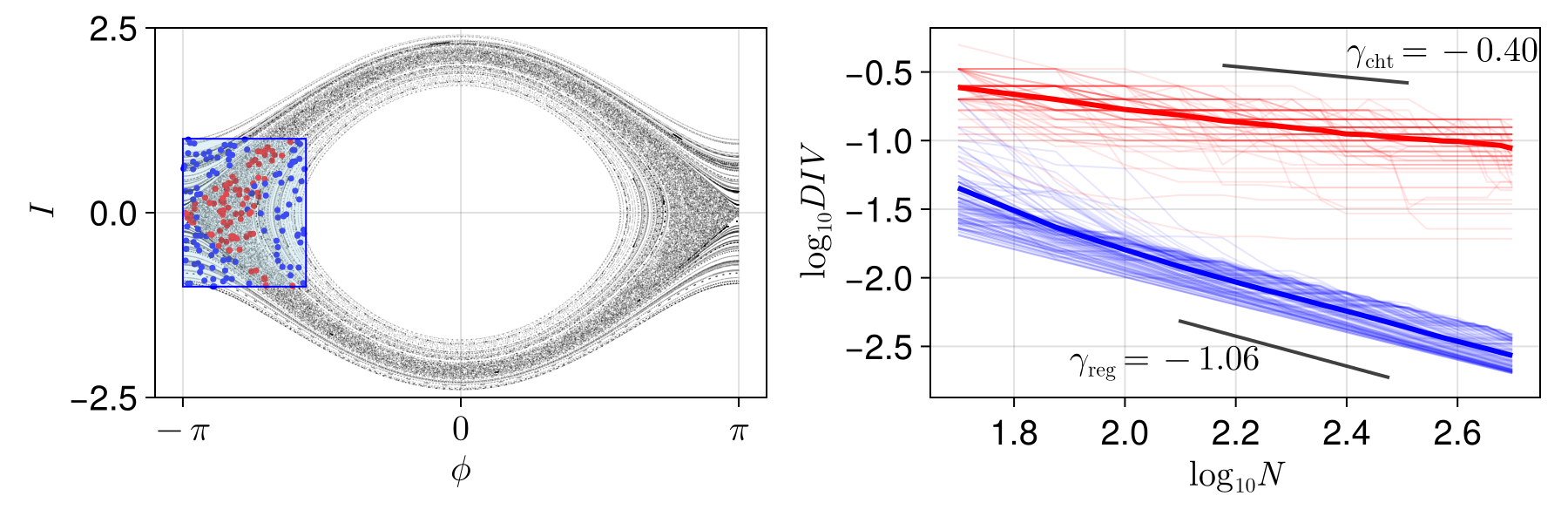}
		\caption{
			(Left) Phase portrait of the modulated pendulum. The light blue box depicts the initial conditions of 200 members of the ensemble used to estimate the power law of $DIV$ vs $N$. The uniform distribution in the highlighted domain provides roughly the same number of regular and chaotic orbits. (Right) Time dependence of measure $DIV$ for chaotic (red) and regular (blue) parts of the ensemble. Initial conditions of trajectories are shown in left panel.}
		\label{fig:figA1}
	\end{figure}

	We consider the model 
	\begin{align}
		\mathcal{J}(I,\phi,t)
		=
		\frac{I^{2}}{2} - 
		(1+\alpha \cos \epsilon t) \cos \phi, \label{eq:modJ}
	\end{align} 
	corresponding to a pendulum model with variable length. 
	\tred{This Hamiltonian is discussed in \cite{aMo22} (chapter $9$).}
	For this model, $T=2\pi/\epsilon$ which is large when $\epsilon \ll 1$. 
	\tred{
		As the time variable $\tau=\epsilon t$ is slow ($\dot{\tau} =\epsilon$, $\epsilon \ll 1$), this model is paradigmatic of slow chaos where $3$ resonances are $\epsilon$ apart. 
		In fact, using trigonometrical identities, $\mathcal{J}$ might be rewritten as
		\begin{equation}
			\mathcal{J}=
			\frac{I^2}{2}+\epsilon J-
			\big(
			\cos \phi
			+\frac{\alpha}{2} \cos(\phi-\tau)
			+\frac{\alpha}{2} \cos(\phi+\tau)
			\big).
		\end{equation}
		The $3$ harmonics are now clearly apparent. 
		By Hamilton's canonical equations, one sees that the centres of the resonances 
		$\dot{\phi}=0$, $\dot{\phi}-\dot{\tau}$
		and $\dot{\phi}+\dot{\tau}$ correspond respectively to the actions values $I=0$, $I=\epsilon$, $I=-\epsilon$. 
	}
	We selected $\alpha=0.25$ and $\epsilon=0.1$. 
	The phase space of $P$ together with the temporal laws of $DIV$ on regular and chaotic components are shown in Fig.\,\ref{fig:figA1}. We found $\gamma_{\textrm{reg}.}=-1.06$ and $\gamma_{\textrm{cht}.}=-0.40$.\\
	
	\tred{The exponents found when starting from the knowledge of observables, including the reconstruction of the phase space or not, are presented in Tab.\,\ref{tab:modJ}.}
	
	\begin{table}
		\begin{tabular}{l|c|c}
			& $\gamma_{\textrm{reg.}}$ & $\gamma_{\textrm{cht.}}$\\
			\hline
			\hline 
			Original $2$D phase space & -1.06 & -0.40 \\
			Observable $I$, reconstruction & -1.08 & -0.40\\
			Observable $\phi$, reconstruction & -1.06 & -0.36 \\
			Observable $I$, no reconstruction& -1.15 & -0.41\\
			Observable $\phi$, no reconstruction & -1.06 & -0.43
		\end{tabular}
		\caption{
			\label{tab:modJ}
			Same as Tab.\,\ref{tab:modK} for the Hamiltonian model of Eq.\,\ref{eq:modJ}.}
	\end{table}

\newpage

\bibliographystyle{apalike} 
\bibliography{biblio}

\end{document}